\documentclass[letterpaper]{raa}            

\usepackage{graphicx,times}             
\usepackage{natbib}
\usepackage{color}
\usepackage{amssymb,amsmath}
\newcommand{\red}{}  
\newcommand{\blue}{}

\begin{document}

   \title{Large solar energetic particle event that occurred on $2012$~March~$7$ and its VDA analysis
}

   \volnopage{Vol.0 (200x) No.0, 000--000}      
   \setcounter{page}{1}          

   \author{Liu-Guan, Ding
      \inst{1}
   \and Xin-Xin, Cao
       \inst{1}
   \and Zhi-Wei, Wang
       \inst{1}
   \and Gui-Ming, Le
       \inst{2}
   }

   \institute{School of Physics and Optoelectronic Engineering, Institute of Space Weather, Nanjing University of Information Science and Technology,
             Nanjing 210044, China; {\it dlgedu@163.com}\\
   \and National Center for Space Weather, China Meteorological Administration, Beijing, 100081, China \\
   }

   \date{Received~~2016 month day; accepted~~2016~~month day}


\abstract{ On 2012 March 7, the STEREO Ahead and Behind spacecraft, along with the near-earth spacecraft (e.g. SOHO, Wind)
situated between the two STEREO spacecraft, observed an extremely large global solar energetic particle (SEP) event in Solar Cycle 24.
Two successive coronal mass ejections (CMEs) have been detected close in time. From the multi-point in-situ
observations, it can be found that this SEP event was caused by the first CME, and the second one
was not involved. Using the velocity dispersion analysis (VDA), we find that for well magnetically connected point, the
energetic protons and electrons are released nearly at the same time. The
path lengths to STEREO-B(STB) of protons and electrons have distinct difference and deviate remarkably from the nominal Parker spiral path length, which is likely due to the
 presence of interplanetary magnetic structures situated between the source and the STB.
Also the VDA method seems only to obtain reasonable results at
 well-connected locations and the inferred energetic particles release times in different energy channels
are similar.
We suggest that good-connection is crucial for obtaining both accurate release time
and path length simultaneously, agreeing with the modeling result of \red{Wang \& Qin (2015)}.
}

   \authorrunning{L.G. Ding, et al}            
   \titlerunning{SEP on 2012 March 7 and VDA analysis}         

   \maketitle

%
%
\section{Introduction}           
\label{sect:intro}
Solar energetic particles (SEPs) are charged particles with energies much greater than those of the bulk solar wind, and
originate from explosive processes at the Sun such as flares and coronal mass ejections (CMEs), termed as ``impulsive''
event and ``gradual'' event respectively.  Historical studies show that
large gradual SEP events are almost always associated with coronal mass ejections (CMEs)
\citep{Kahler.etal84,Reames95,Kahler96,Gopalswamy.etal02,Cliver.etal04,Tylka.etal03,Kahler.Vourlidas13}.
However, not all fast and wide CMEs lead to large SEP events \citep{Kahler96,Ding.etal13,Ding.etal14}.
\citet{Kahler96,Kahler.etal00} noted that although the the maximum energy and intensity of energetic particles in large
SEPs events are generally correlated with CME speed, the scatter is very large, and suggested that the ambient superthermal
seed particles can be another important factor for producing large SEP events. The seed population commonly derive from
the materials of solar flares \citep{Mason.etal99,Mason.etal00,Cane.etal06,Ding.etal15} and the preceding CMEs
\citep{Gopalswamy.etal02,Gopalswamy.etal04,Li.etal12,Ding.etal13,Ding.etal15}.

\citet{Gopalswamy.etal02} first showed that the interaction of two CMEs is an important aspect of SEP production, and the shocks
can preferrentially accelerate particles from the material of the preceding CMEs rather than from the quiet solar wind.
Later, \citet{Gopalswamy.etal04} found that there is a strong correlation between high particle intensity events and the
existence of preceding CMEs within $24$ hrs ahead of the primary CMEs, but poor correlation with the flare class.
Then, \citet{Li.Zank05a} proposed that two consecutive CMEs may provide a favorable environment for particle acceleration.
Recently, \citet{Ding.etal13} suggested that
large gradual solar energetic particle events are often associated with twin-CMEs, a  scenario first proposed by
\citet{Li.etal12}. In this scenario, the preceding CME can provide enough enhanced
turbulence and seed population ahead the main CME-driven shock to generate a more efficient particle acceleration process
compared to single CME \citep{Li.etal12,Ding.etal15}. Based on a statistical analysis of the SEP events in Solar Cycle 23,
\citet{Ding.etal13} found that 61\% twin CMEs lead to large SEP events as compared to
only 29\% single fast CMEs leading to large SEP events.
However, not all twin-CMEs lead to a large SEP event \citep{Ding.etal13,Ding.etal14}, and the relevance of
CME-CME interactions for larger SEP events remains unclear \citep{Kahler.Vourlidas14}.

After the launch of STEREO mission, CME observations from multiple vintage points became available. By using the co-observations
of STEREO combined with SOHO and SDO, two CMEs were identified to erupt successively from the same active region within merely
3 minutes in the 2012 May 17 GLE event \citep{Shen.etal13}.
Also, \citet{Ding.etal14a} investigated the eruption
and interaction of two CMEs during the large SEP event occurred on 2013 May 22 using multiple spacecraft observations,
and found that the release times of proton and electron agreed with the time when the second CME caught up with the
trailing edge of the first CME, indicating that CME-CME interaction (or shock-CME interaction) plays an important
role in the process of particle acceleration in that event.

The time at which SEPs are first released into interplanetary space, and its relation to CMEs and various photon emissions,
are important clues to the site and nature of the SEP acceleration mechanism
\citep{Lin.etal81,Kahler94,Tylka.etal03,Reames09a,Lin11}. The velocity dispersion analysis (VDA) is an often used tool for
obtaining the solar particle release (SPR) time in studying impulsive and gradual SEP events.
Plotting onset times versus $v^{-1}$ yields a line with the initial SPR time in the solar vicinity as the intercept and
the magnetic path length as the slope.
This SPR time defines only the earliest acceleration or release of particles that are unscattered in transit.
In this practice, SEPs of different energies are assumed to be released at the same time and the same location near the Sun.
Using this method, \citet{Tylka.etal03} examined two impulsive events and three GLEs and found that the SPR times in
impulsive events occurred precisely at the peak time of hard X-rays, while that in GLEs often coincided with CME-driven
shocks. \citet{Reames09a} tested all the GLE events in Solar Cycle 23 with the VDA method and found that the path length of
GLEs vary from 1.1AU to 2.2 AU and the SPR times in all of the GLEs occur after the onset of the shockwave-induced type II
radio emission. Further, \citet{Reames09b} found that acceleration for magnetically well-connected large GLEs begins at
$\sim2$ solar radii, in contrast to non-GLEs that have been found to be strongly associated with shocks above 3 solar radii,
as well as \citet{Gopalswamy.etal12}.
From the VDA analysis, \citet{Tan.etal13} noted that the deduced path length of low-energy electrons from their release site
near the Sun to the 1 AU observer is consistent with the ion path length deduced by \citet{Reames09a,Reames09b}.

However, for many gradual SEP events, the release of particles accelerated to different energy by CME-driven shock becomes
complicated and doesn't occur at the same time \citep[see e.g.][]{Li.etal03,Li.etal05,Ding.etal14a}.
Recently an alternative VDA practice is used where the path length of energetic particle is assumed to be the nominal
Parker spiral and then obtain the release times for particles of different energies. Using this method,
\citet{Kim.etal14} suggested a new classification scheme of SEP events based on the release timing relative to flares and
energy-dependent flux enhancement, unlike the conventional classification of SEP based on whether the flux time profile is
impulsive or gradual. And \citet{Kim.etal15} also implied that there were some different characteristics of different
groups with different origins and acceleration processes.
Following this practice \citep{Kim.etal14,Kim.etal15}, \citet{Ding.etal16} found that there
were two particle release processes that occurred in the 2012 May 17 GLE event: the first one is consistent
with particles accelerated at the solar flare, and the second one is consistent with particles accelerated at the
associated CME-driven shock.

Obviously, the inferred release times and the path length have some uncertainties in the VDA practices.
\citet{Dalla.etal03} found that the derived path length at Ulysses are 1.06 to 2.45 times the length of a Parker spiral magnetic field line connecting the spacecraft to the Sun, and the time of particle release from the Sun is between 100 and 350 min later than the release time derived from in-ecliptic measurements.
To apply the VDA method more reasonably, many numerical simulation studies have been done to investigate the validity of the method. Only when the parallel mean-free path (MFP) are large enough ($\lambda_{||}>0.3$AU), the interplanetary scattering can have an insignificant effect on the derived solar release time \citep{Lintunen.Vainio04}. Further, when the background level is below 0.01 of the peak intensity of the flux, the onset time of the SEP event can be identified precisely \citep{Saiz.etal05}. Recently, \citet{Wang.Qin15} analyzed the accuracy of the VDA method and proposed that an ideal SEP event for VDA analysis should meet the following conditions: impulsive source duration, large parallel MFP, low background level, and good connection between the observer and the source.


To argue whether twin CMEs are the cause of large SEP events, the release time of SEPs at the Sun has to be later than
the eruption of the second CME.  In principle, one can use the VDA method to obtain the solar release time of SEPs and examine
if this is after the eruption of the second CME. However, the VDA methods are subject to uncertainties. Therefore for twin CMEs erupting closely in time (e.g., within an hour), it is hard to obtain the correct ordering of the SEP release time and
the eruption time of the second CME unless accurate VDA can be performed.

In this work, we performed a case study and examined a SEP event where two consecutive CMEs were found.
The event occurred on 2012 March 7 and was observed by both STEREO-A and -B, WIND and ACE.  Applying the VDA to multi-spacecraft
observations allows us to estimate the SEP release time more accurately than that to a single spacecraft observation. We found that
the VDA analysis is best applied when the spacecraft is magnetically well-connected to the acceleration site.

\section{Observations}
\label{sect:obs}

   \begin{figure}[htb]
   \centering
   \includegraphics[width=0.4\textwidth, angle=0]{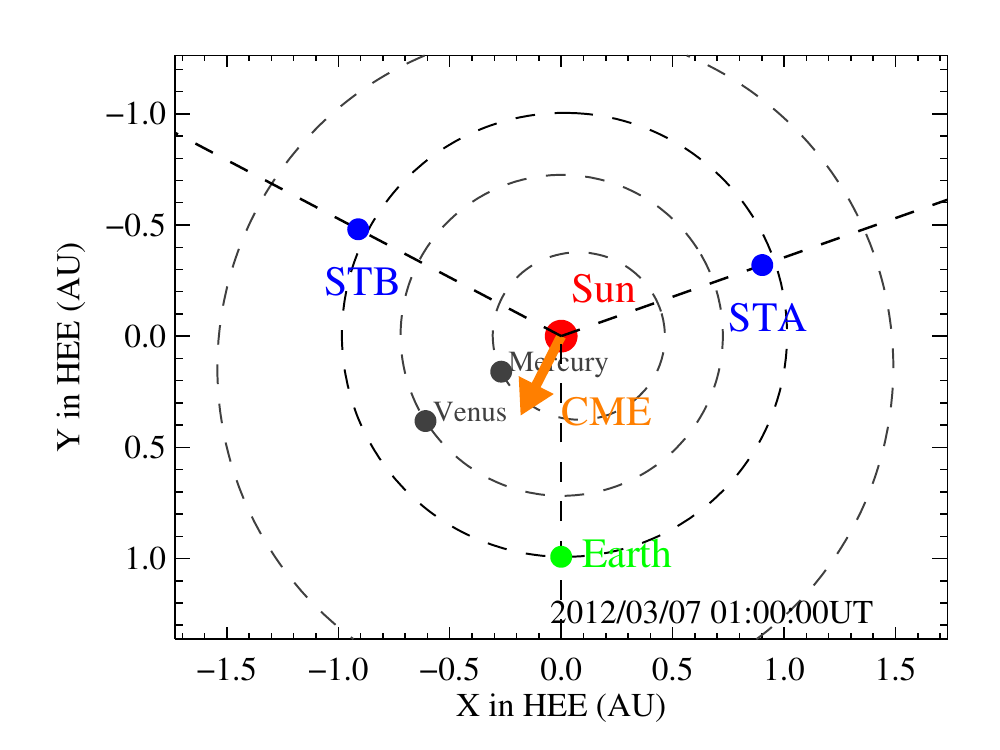}
   \caption{The positions of the STEREO-A/B spacecraft and the Earth. The orange arrow indicates the propagation direction of the associated CME.}
   \label{Fig1}
   \end{figure}

We use in-situ and remote-sense observations from multiple-vantage spacecraft in this study. They included energetic
particle fluxes detected from the STEREO/HET \citep{Rosenvinge.etal08} and similar observations from near-Earth
spacecraft (e.g. SOHO/ERNE \citep{Torsti.etal95,Valtonen.etal97} and Wind/3DP \citep{Bougeret.etal95,Lin.etal95}), coronagraph
observations from the STEREO/SECCHI \citep{Howard.etal08} and the SOHO/LASCO \citep{Brueckner.etal95}, and radio observations from
the STEREO/WAVES \citep{Cecconi.etal08} and Wind/WAVES spacecraft respectively.

Figure~\ref{Fig1} shows the relative configuration of three spacecraft at 01:00UT on 7 March 2012. The angular separation
between the Earth and STEREO-A is $109.5^\circ$; the angular separation between the Earth and STEREO-B is $117.8^\circ$.
Since the associated AR (No.11429 in NOAA) is located at about E20 from the Earth, the event was a backside event for STEREO-A
 and a western limb event for STEREO-B. The propagation direction of two successive CMEs is also marked by the orange arrow.

\subsection{CME observation}
\label{sub:cme}

   \begin{figure}[htb]
   \centering
   \includegraphics[width=0.6\textwidth, angle=0]{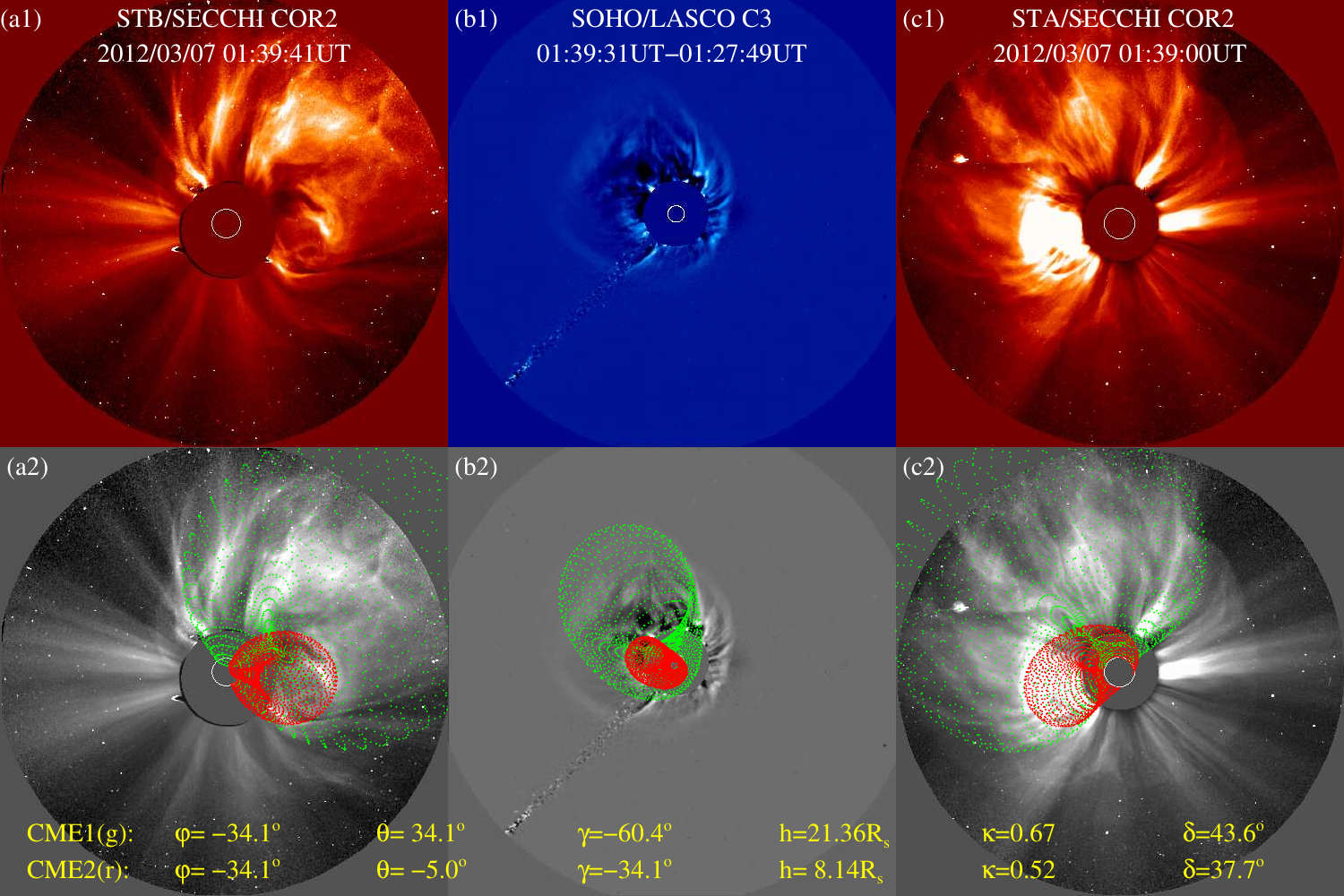}
   \caption{The two CMEs during this SEP event detected by STEREO-A(/B)/SECCHI and SOHO/LASCO, and their GCS model fitting results. }
   \label{Fig2}
   \end{figure}

On 2012 March 7, an X5.4 flare (onset: 00:02UT, peak: 00:24UT, end: 00:40UT) at E$27^\circ$ was followed by an X1.3 flare
(onset: 01:05UT, peak: 01:14UT, end: 01:23UT) at E$17^\circ$. Two fast halo LASCO CMEs were accompanied with these events,
the first observed above the occulting disk at 00:24UT (with a speed of 2684 km/s
\footnote{http://cdaw.gsfc.nasa.gov/CME\_list/UNIVERSAL/2012\_03/univ2012\_03.html}), and the second at \red{01:36UT} (1825 km/s$^1$).

Coronagraph observations made by SOHO/LASCO, STEREO-A(STA)/COR2, and STEREO-B(STB)/COR2 are shown in Figure~\ref{Fig2}.
Panel a1(c1) is the STB(STA) COR2 image at 01:39UT. Panel b1 is the running difference of the SOHO/LASCO C3 image
(01:39UT - 01:27UT). The envelopes of the two CMEs can be clearly seen from these panels. To better examine these two CMEs,
we also used the Graduated Cylindrical Shell (GCS, \citet{Thernisien.etal06,Thernisien.etal09}) model to fit the CME1 and CME2. In bottom panels of Figure~\ref{Fig2}, the green grids denote the
fitting result of CME1, and the red grids denote the fitting result of CME2. From the fitting results, one can see
clearly that the propagation directions of these two CMEs have  similar longitudes but distinct latitudes in space.

\subsection{Radio bursts}
\label{sub:radio}

   \begin{figure}[htb]
   \centering
   \includegraphics[width=0.5\textwidth, angle=0]{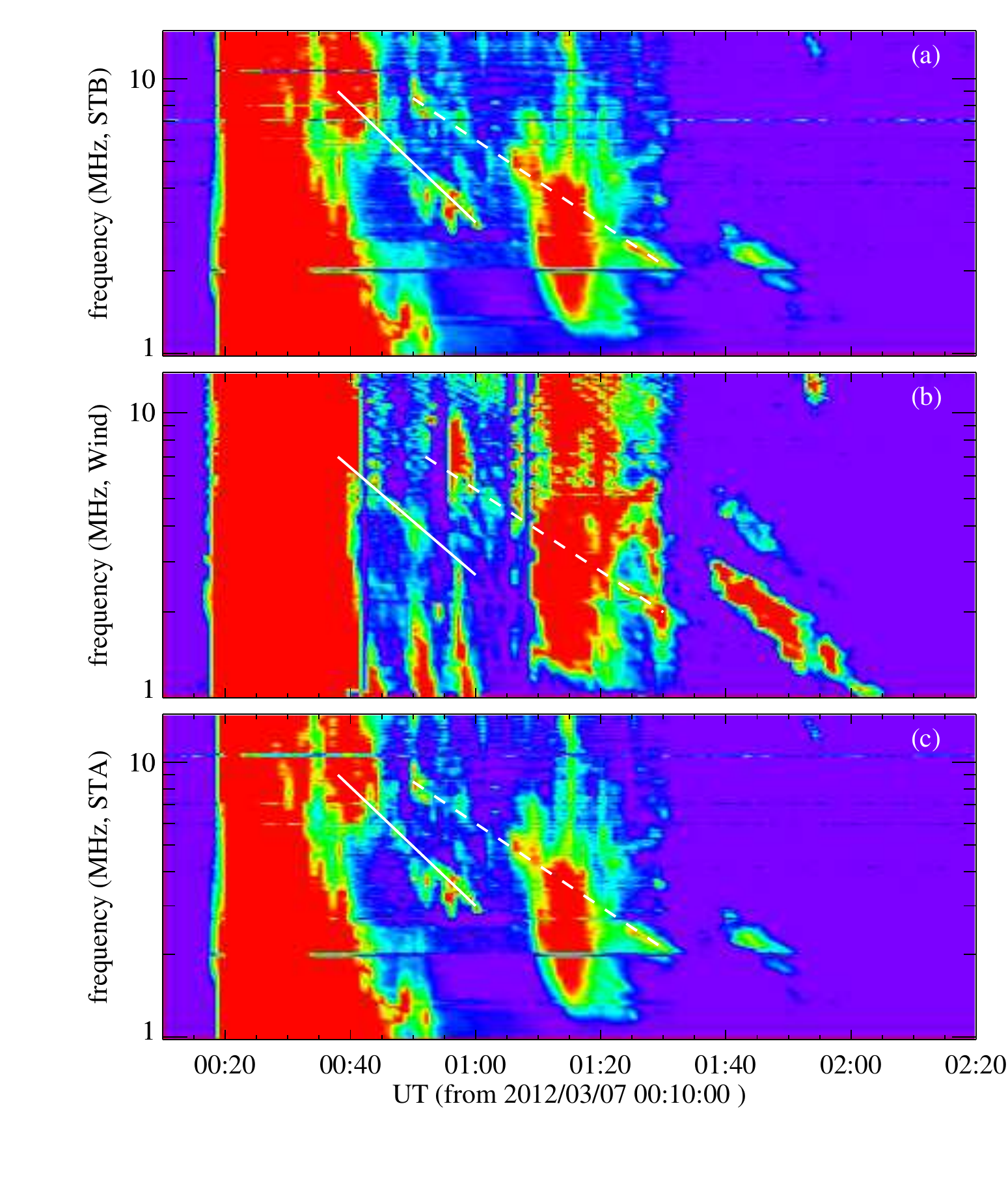}
   \caption{The radio bursts observations detected by three spacecraft (STEREO-A(B)/WAVES and Wind/WAVES). The white line shows the shift of type II radio bursts. }
   \label{Fig3}
   \end{figure}

Type II radio bursts are caused by shock-accelerated electrons radiating at local plasma frequencies.
As the shock propagates out, the ambient plasma density drops and the radio burst drifts to lower frequencies.
Type II radio bursts have been used as a diagnostic of the CME and its driven shock in the corona and interplanetary space
in studying SEP events \citep[see e.g.][]{kahler82,Cane.etal02,Cliver.etal04,Gopalswamy05}. Figure~\ref{Fig3} shows
radio observations in the frequency range of 1-14MHz detected by the WAVES instrument onboard STEREO and Wind respectively.
Multiple episodes of type II and type III radio emissions can be seen from the radio dynamic spectra. All three spacecraft
observed a bright type III radio burst beginning at about 00:18UT, which we intepret to be associated with the first flare
and coronal eruption. A type II radio burst followed from about 00:35UT to 01:00UT. It was not continuous and is marked
by white solid lines in each panel. Often both plasma emissions at the fundamental and the second harmonic frequencies
can be observed. We believe that this is also the case in this event. The second harmonic branch can be clearly identified
in STB and STA, marked by the white dash lines in the panel (a-c). This type II radio burst is evidently associated with
the shock driven by the first CME.

Later, the second type III radio burst was detected from about 01:09UT to 01:30UT. It is associated with the second
flare eruption. There are also other type II radio burst episodes shown between 01:38UT and 02:05UT. From the radio spectra
of the Wind/WAVES shown in the panel (b), we can clearly see the fundamental and second harmonic branches. These are clearly
not the continuations of those type II radio emissions associated with the first CME. So we intepret
the second type II radio bursts to be associated with the shock driven by the second CME in this event.
From the start frequency of these two episodes of type II radio bursts, we see  that the first shock is formed at
lower height than the second one.

\subsection{In-situ observation of SEPs}
\label{sub:sep}

   \begin{figure}[htb]
   \centering
   \includegraphics[width=0.5\textwidth, angle=0]{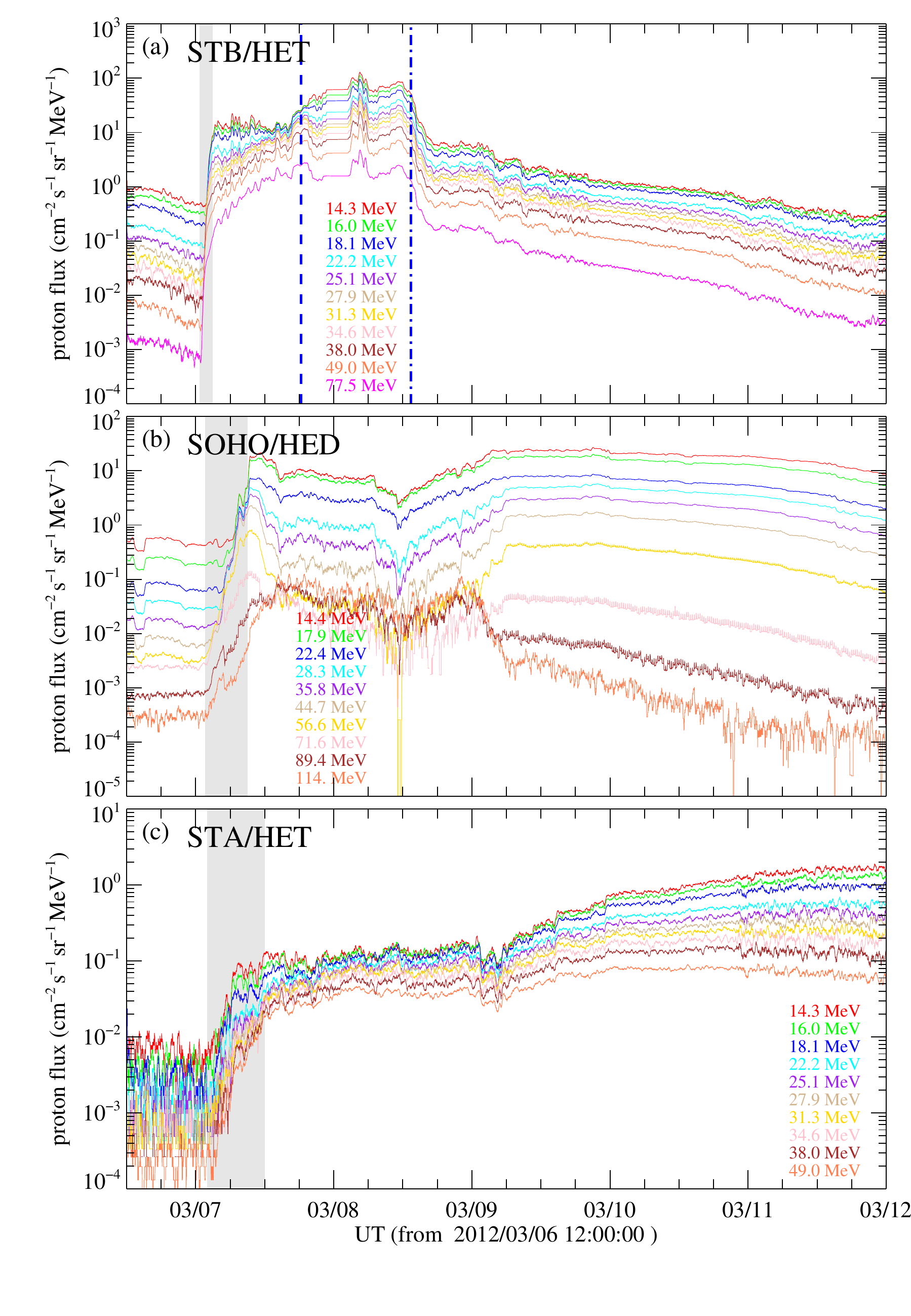}
   \caption{The solar energetic protons detected by three spacecraft (STEREO-A/B and SOHO/ERNE). The shaded areas denote
the interval of energetic proton enhancement in each observation.
The vertical dash line and dot-dash line in the panel (a) indicate the time of IP shocks respectively (solar wind observations shown in Figure~\ref{Fig6}).}
   \label{Fig4}
   \end{figure}

   \begin{figure}[htb]
   \centering
   \includegraphics[width=0.5\textwidth, angle=0]{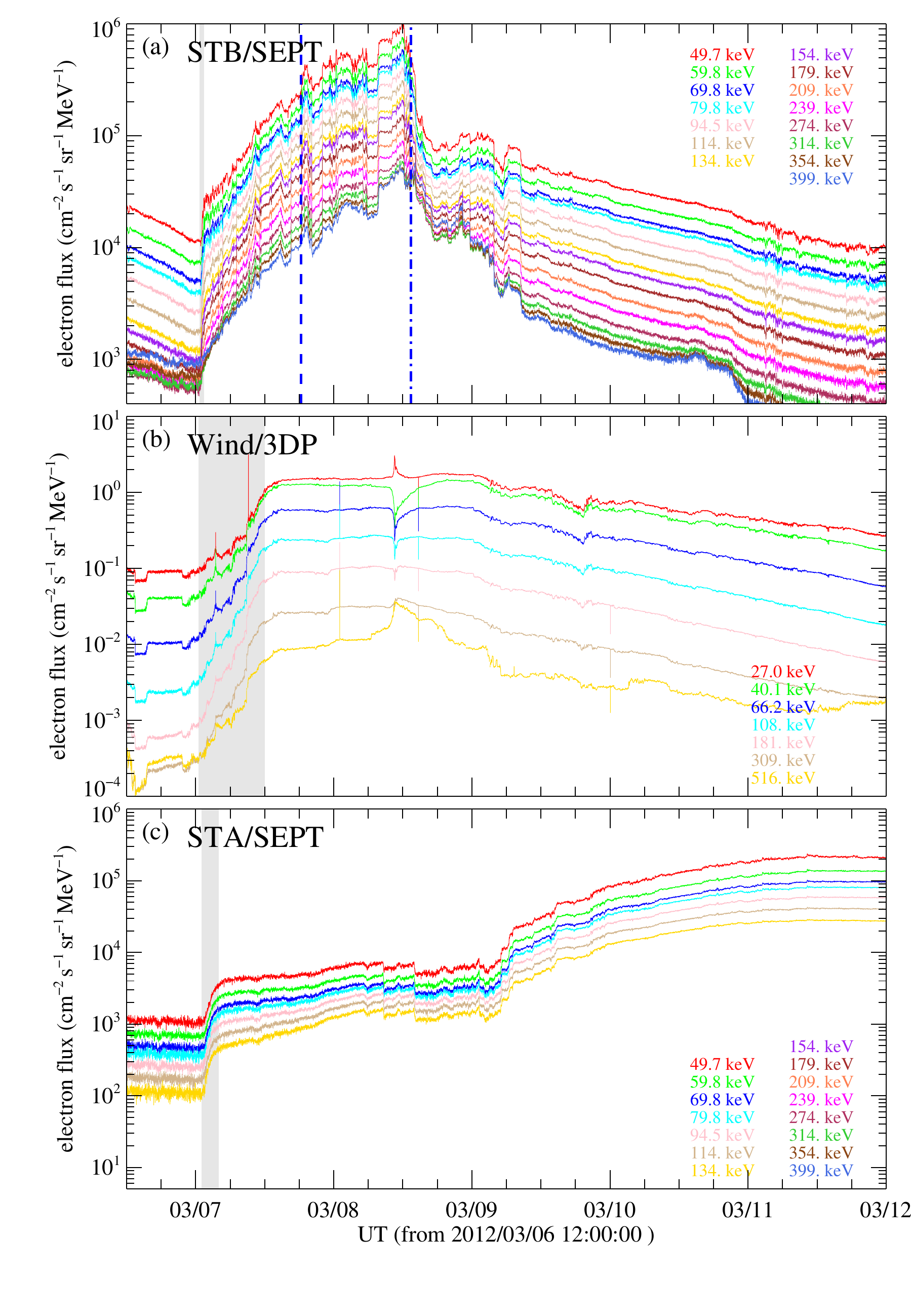}
   \caption{The in-situ observations of solar energetic electrons detected by three spacecraft (STEREO-A/B and Wind/3DP).
The shaded areas denote the intervals of energetic electron enhancement in each observation.
The vertical lines in the panel (a) show the time of IP shocks, the same as those in Figure~\ref{Fig4}(a).}
   \label{Fig5}
   \end{figure}

Figure~\ref{Fig4} shows one-minute averaged energetic proton observations from the STB(STA)/HET and SOHO ERNE/HED
respectively in panel (a), (b) and (c) during the onset of this SEP event. The shaded areas indicate the ascending period
in each in-situ observation. As shown in the figure, the event was a global event and seen in all three spacecraft.
 The HET  instrument onboard STB was  best magnetically connected. It detected prompt enhancements in all energy channels from
$\sim$01:15UT to $\sim$02:10UT on March 7, with clear velocity dispersion. These onset times are delayed
by about 40 minutes (due to propagation) from the release near the sun (see section~\ref{sect:vda}).
These SEPs are clearly consistent with the first solar eruption before 01:00UT and inconsistent with the second solar
eruption after 01:00UT.
Then SOHO detected gradual enhancements in all energy channels within $4.5$ hours after about 02:10UT. The peak flux intensity
in SOHO was $\sim 10$ times lower than that observed in STB. Later, STA also detected gradual enhancement in all energy channels
within about $3$ hours, and the peak intensity was $\sim$10 times lower than that observed in SOHO too. The onsets
in SOHO and STA were after the second CME.  Therefore from only the SOHO and STA observation, one can not tell
if the in-situ SEPs are due to the shock driven by the 1st CME, or the 2nd CME.

However, with STB observation, which is well magnetically situated,
we can clearly identify that these SEPs are accelerated by the 1st CME-driven shock.
We note that the first shock was formed at a lower height, which is more suitable for leading to high intensity SEP event or
GLE event \citep{Reames09b,Gopalswamy.etal12,Mewaldt.etal12}.

Our study shows that multi-vantage-point observations are crucial in understanding SEP events.

%
%

We next examine the observations of energetic electron from multi-vantage points in Figure~\ref{Fig5}.
From the panel (a) of this figure, the electron intensity at STB/SEPT clearly increased rapidly in all energy channels
by $\sim$00:50UT. About 30 minutes later, the STA detected gradual enhancements in all energy channels shown in the panel (c).
Unexpectedly, the energetic electron flux detected by Wind/3DP seem to start to increase very gradually at about 00:00UT
as shown in the panel (b).
\red{
Due to the passage of an IP shock through the Wind spacecraft at about 03:30UT on March 7, which may cause a slow rise beginning several hours before the shock \citep{Tsurutani.Lin85},
the time profiles in that time are contaminated and it is difficult to accurately identify their onsets.
}
Again, only STB, the best-connected detector, provides unambiguous association between the energetic electrons and
the first CME-driven shock. Neither STA nor SOHO can reveal this association.

\subsection{In-situ observation of solar wind}
\label{sub:sw}

   \begin{figure}[htb]
   \centering
   \includegraphics[width=0.5\textwidth, angle=0]{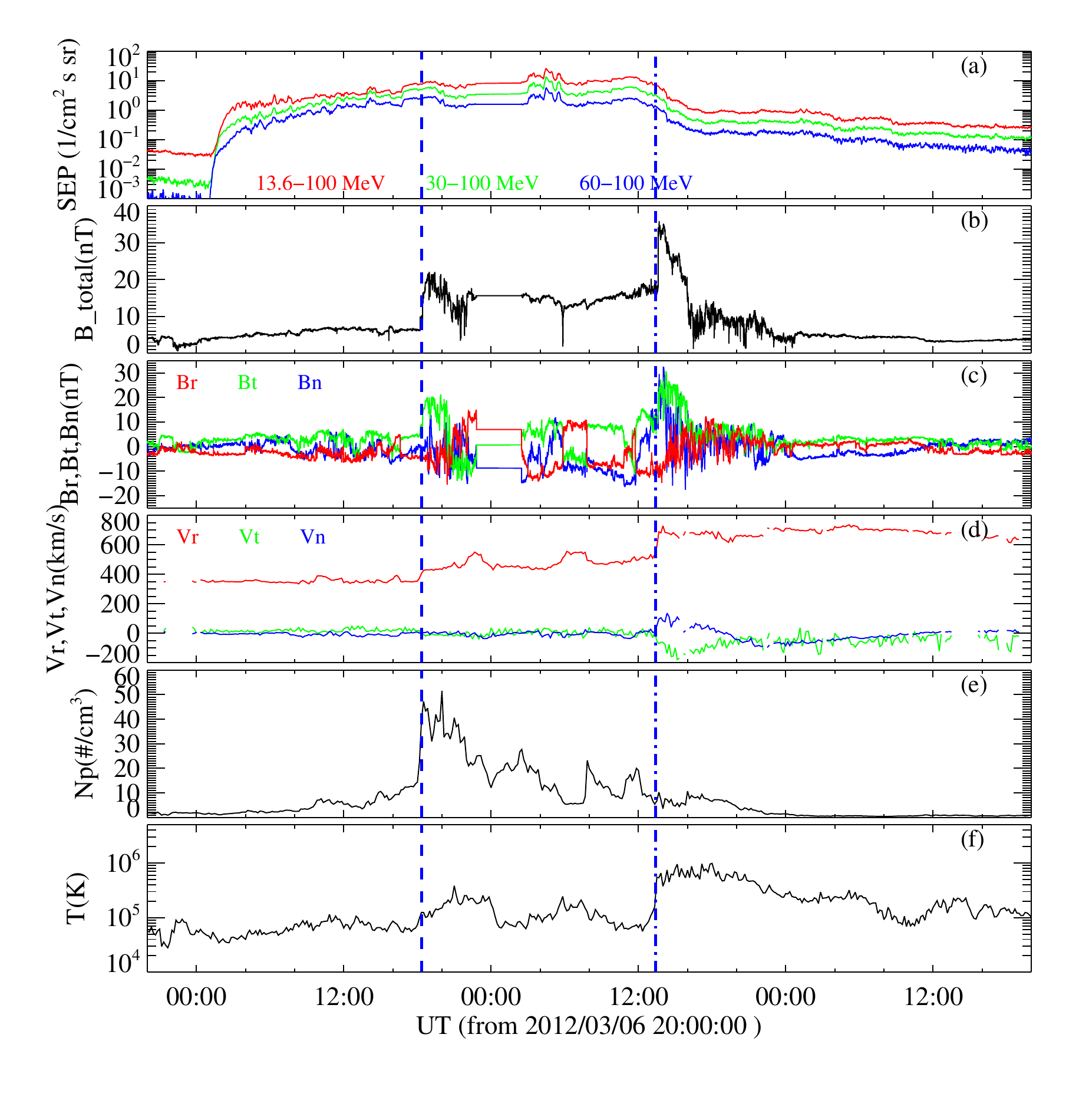}
   \caption{The solar wind observations in the point of STEREO-B.
   The vertical dash lines and dot-dash lines in each panel represent two IP shocks.}
   \label{Fig6}
   \end{figure}


The intensity time profile of solar wind detected by STB became complicated by the passage of shocks, sheath,
and interplanetary (IP) coronal mass ejection (ICME) on March 7-8.
\red{
In particular, the intensities weakly increased at about 18:20UT on March 7 and sharply decreased at about 13:25UT on March 8, which were probably associated with the passage of two ICMEs and/or their shocks respectively as shown in Figure~\ref{Fig6}. Increases of SEP intensities observed in association with the passage of transient interplanetary (IP) shocks are known as energetic storm particle (ESP) events \citep{Bryant.etal62,Reames99}. There are also cases where there are decreases in the time profiles associated with shock passage. For example,  \citet{Hietala.etal11} noted that such decreases  may occur due to accelerated particle trapped between shocks (e.g. see the Wind observations at the second shock passage in \citet{Hietala.etal11}).
}
The first shock (denoted by vertical dash lines) followed by
a hypothetical ICME structure was detected at about 18:20UT, which erupted at March 5 from the AR 11429. Because this ICME is
situated between the source and STB when the accelerated particles start to escape in the solar vicinity, the propagation path
length of particles observed by STB presents a distinguishable difference from the nominal Parker spiral path length corresponding
to solar wind speed (see VDA results below). The second shock (denoted by vertical dot-dash lines) followed by an ICME detected at about
13:25UT on March 8 was probably produced by the first CME launched at 00:24UT on March 7 mentioned before. This shock can also be observed
by the spacecraft near the Earth (not shown here).
\blue{
For the energetic protons and electrons, the intensities in STB and SOHO present a nominal decay phase after the passage of the shock, while those in STA show a second enhancement in this period (after $\sim$05:00UT on March 9). It's very interesting to find that there is a corotating interaction region (CIR) passing through the STA coincidentally at that time, which may be considered to contribute to the second enhancement, as the CIR is also an efficient particle accelerator \citep{Fisk.Lee81,Reames.etal97,Zhao.etal16}.
}

\section{Velocity Dispersion Analysis}
\label{sect:vda}

   \begin{figure}[htb]
   \centering
   \includegraphics[width=0.5\textwidth, angle=0]{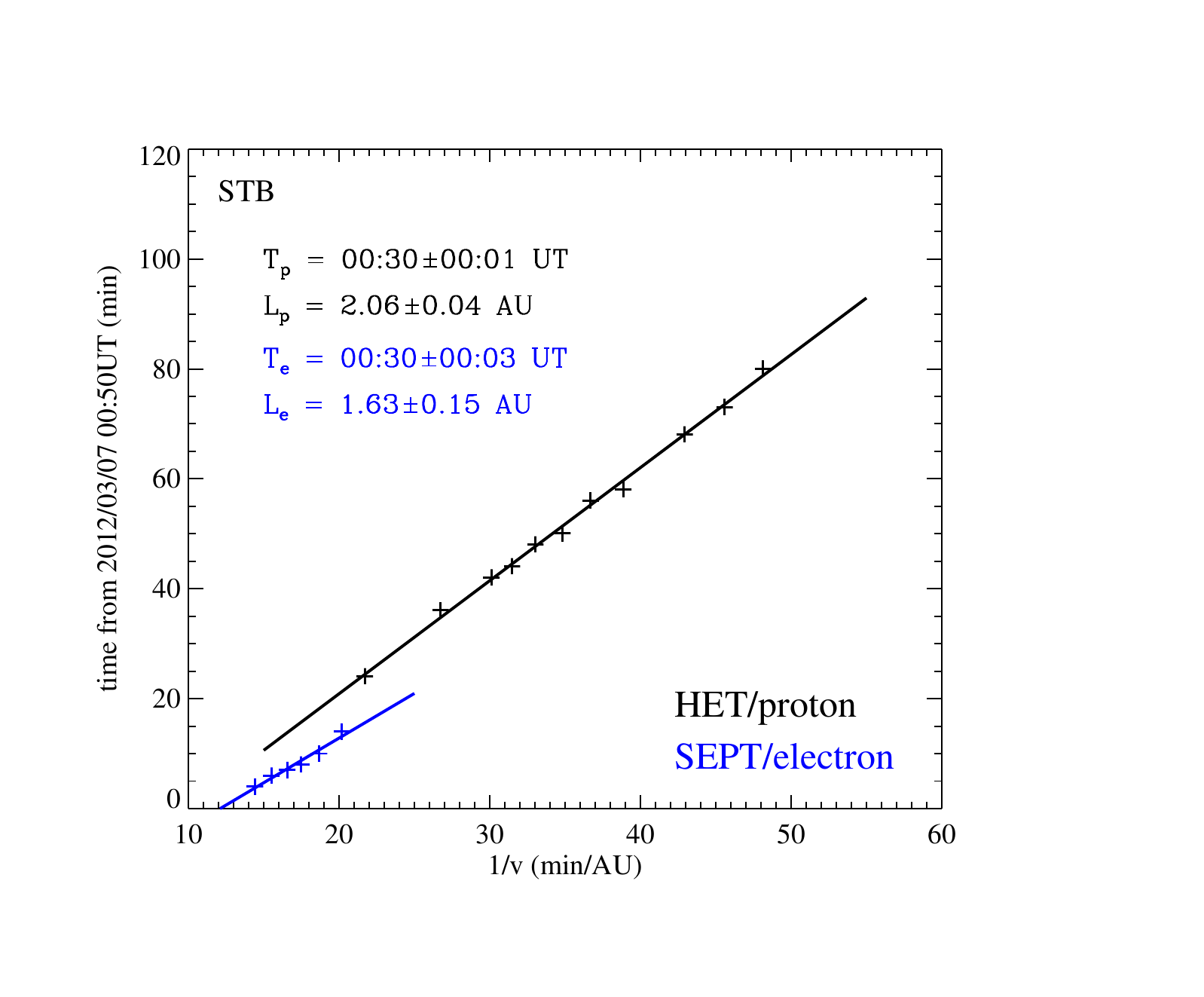}
   \caption{The velocity dispersion analysis (VDA) of energetic protons (black cross line) and energetic electrons
in low energy channels (blue cross line) detected by STEREO-B. }
   \label{Fig7}
   \end{figure}

   \begin{figure}[htb]
   \centering
   \includegraphics[width=0.5\textwidth, angle=0]{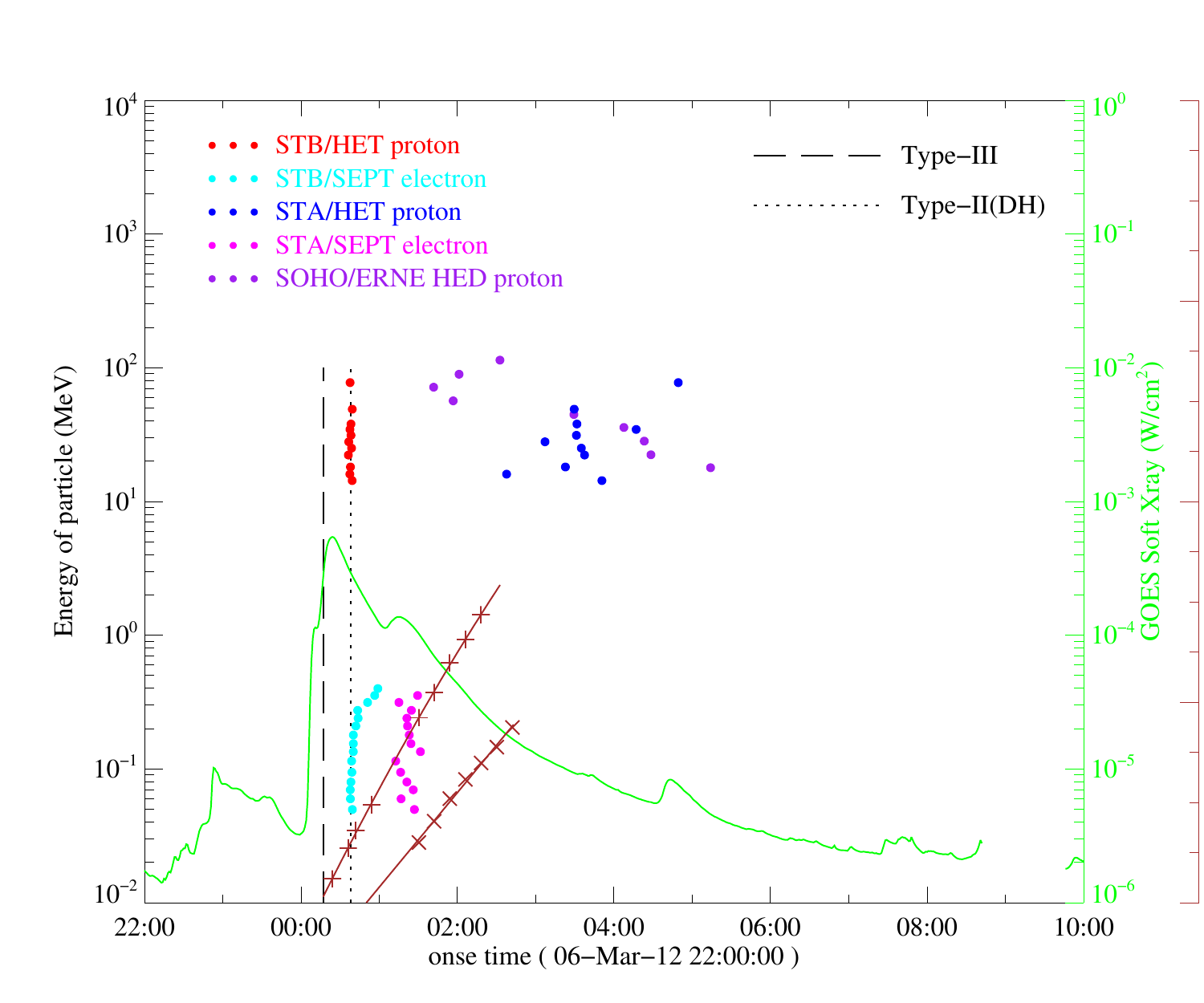}
   \caption{Solar particle release times near the sun of SEPs detected by different spacecraft in different energies and
related eruptive phenomena including flare, CME and radio bursts. The green line shows the soft X-rays detected by GOES
in $1-8\AA$. The brown line with crosses displays the heliocentric height of CME as a function of time. }
   \label{Fig8}
   \end{figure}

To compare the solar energetic particle releases with those of other associated solar activities, it is necessary to compensate
proton travel time from the Sun corresponding to their energies. We perform the VDA of the proton data obtained from STEREO/HET
and SOHO ERNE/HED, and the electron data obtained from STEREO/SEPT and Wind/3DP. To obtain the solar release time, we use
\begin{equation}
 t_{onset}=t_{rel}+t_{trans}=t_{rel}+L/v(E)
 \label{eq1}
\end{equation}
where $t_{rel}$ is the particle release time at its source region near the Sun; $t_{trans}=L/v(E) $ is the travel time for the
particle with energy $E$; $L$ is the propagation path length and $v(E)$ is the velocity of energetic particle with energy $E$.
\red{
We determined the onset times by following the procedure outlined in \citet{Ding.etal14a,Ding.etal16}. The onset time is decided by $f(t_{onset})=<f>+2\sigma$, where $<f>$ is the average of the pre-event background and $\sigma$ is its standard deviation. So, the onset time signals the time when the distribution function $f$ becomes $2\sigma$ higher than the background. The uncertainty of the onset time mainly comes from possible measurement uncertainty, such as data resolution.  We also calculate a possible uncertainty by using either $3\sigma$ or $1\sigma$ \citep[e.g. see][]{Ding.etal16}.
}

For the data of energetic proton and electron obtained from STEREO and near Earth spacecraft (SOHO and Wind), we first performed
the VDA assuming all first arriving energetic particles are released at the same time from their source and
propagating with scatter free along the IP magnetic field lines.
 \red{
It is hard to address the uncertainty due to the scatter-free assumption, because the pitch angles of the first arriving energetic particles can not be determined exactly along the transport path from the Sun to the the observer. However, if we use Equation~(\ref{eq1}), a nonzero pitch angle would result in the experimental path length being larger than the actual path length by approximately a factor of $1/cos\theta$, with $\theta$ representing the pitch angle of the first arriving particles, and estimates of the release time are accurate to the order of several minutes for the first practice of VDA \citep{Dalla.etal03,Saiz.etal05}.
 }

 Unfortunately, except STB (the best-connected spacecraft),
all other observations show  poor velocity dispersion and the inferred VDA results are unreliable.  Here we only display the results
of the STB observations. As shown in Figure~\ref{Fig7}, the $t_{onset}s$ are plotted as a function of $1/v$ for energetic protons and
electrons. The black cross points denote the protons in all energy channels observed by HET and the blue cross points denote the
electrons in low energy channels observed by SEPT. It is obviously indicated that the velocity dispersion distribution of
the protons does not overlap with that of the electrons. We fitted the points of energetic protons and electrons separately.
From the fitting results, it is interesting to find that the protons and electrons were released at the same time $\sim00:30$UT,
but they have different propagation path lengths. The path length of protons is $\sim2.06$AU, and that of electrons is $\sim1.63$AU,
while the nominal Parker spiral length from the Sun to STB with a solar wind speed of 353km/s is $\sim 1.25$AU. This
path length of energetic particles may be caused by the presence of the ICME structure situated between the source region and the
detector.  Such a structure may be also the reason that the path length of protons are distinct from that of electrons.

In order to reveal the release process of energetic particles at different longitudes, the release times in all energy channels
derived from the data of different detectors are shown in Figure~\ref{Fig8}. Here we assume that the particles in different
energies transport along the nominal Parker spiral if there is no special magnetic structure situated between the Sun and the
detector. To compare with, e.g., CME eruption and radio observations, we add 8.3 min, the travel time of light from the Sun to
the Earth, to the release time $t_{rel}$. For the energetic protons observed by STB, we use the deduced path length 2.06AU
obtained above as their actual transport path length, as well as 1.63AU as that of energetic electrons. Under the assumption of
scatter free, the release times of protons and electrons in different energy channels are obtained, as indicated by the red points
and cyan points in the figure respectively. From the distribution of the release times versus particle energies, it is clear that
all protons and electrons  (except from high energy channels) observed by STB were released at about the same time.  This time also
coincided with the start time of DH type II radio bursts (denoted by the vertical dot line) that is associated with the first
 CME driven shock.
The simultaneity of the particle release in all energy channels for proton and low energy electrons supports
the proposal that the interplanetary magnetic field in this event was distorted by some IP structures and the
the the path length electrons and protons are different. It also suggests that
the result from the VDA as shown in the Figure~\ref{Fig7} is reliable.

As a comparison, we also calculate the release times of energetic particles detected by STA and SOHO using the VDA method.
We assume nominal Parker spiral path length (1.16AU for STA and 1.18AU for SOHO) as no considerable IP magnetic structures were
between the Sun and STA and SOHO.
From the figure, one can see that the release times of protons in different energy as observed by SOHO and STA have
large scatterings. This imply that energetic protons in different energy channels may not be released at the same time
or that they have propagated along different paths. It should be noted that the SOHO and STA are not well-connected with the
source region in this event. So if we apply the VDA to SOHO or STA in this event (not show the results here), one can find
that it does not yield a fine result of release time or path length like STB.

Since the particle release times for the STA and SOHO show large scatterings, these release times can not be used to
compare with other solar activities. This of course is because both STA and SOHO are  poorly-connected spacecraft.
Indeed, if one were to use the releases of protons observed by SOHO and STA, one may mistakenly arrive at the
conclusion that these SEPs are associated with the shock driven by the second CME.

As shown in the figure, the energetic electrons detected by STA/SEPT seem to be released at almost the same time but with
slight scatter. We also apply the VDA to this data to obtain the electron release time and path length.
It's interesting that the calculated path length is similar to the nominal Parker spiral path length. The resultant release time
is about 01:13UT, which coincided with the peak time of the second flare, but the uncertainty of the result is large \red{($\pm9$ minutes)}.
Note that the protons and electrons observed by STA yield the different release times.

From our event we see that it should be very cautious when using the VDA to obtain the path length and SEP release time.
Under the assumption that particles are released at the same time from the solar vicinity in different energy channels, it seems that
reasonable transport path length from the Sun to the observer can be obtained, even if there is magnetic structures situated between the source and the detector.
However, the release times may be different for different observers located at different longitudes, suggesting that the release time
can strongly depend on if the spacecraft is well magnetically connected or not. This of course is realistic
 since the acceleration site
can connect to different field lines at different times.
Usually, the energetic particles are released later when the foot point of the observer is far away
the source region, because of the shock expansion along the longitude during the propagation outward.
Alternatively, energetic particles may undergo cross-field diffusion.
Finally, we note that there can exist significant differences between ions and electrons in both the release time and the
path length.

\section{Conclusion and Discussion}
\label{sect:discussion}

In this paper, we performed the case study of the SEP event occurred on 2012 March 7 using the in-situ and remote-sense
observations of the multi-vantage spacecraft. There were two X-class solar flares and fast and wide successive CMEs erupting
from the same source region during the event. These two CMEs agree with the proposed twin-CME scenario
\citep{Li.etal12,Ding.etal13,Ding.etal14a}, except that in this event the second CME was slower than the first one.
Analyses of particle onset and release indicate that particle acceleration in this event was associated with the first
solar event/CME and that the second CME was not involved, in agreement with the result of \citet{Richardson.etal14}.
In addition, there is no obvious indication in the STB intensity time profile of a second particle injection that was
associated with the second solar event. That energetic particles are accelerated at the second CME-driven
shock, as suggested by  the twin-CMEs scenario \citep{Li.etal12}, may not be in operation in this event.

The release time of the protons and electrons detected by the best-connected STB coincided well with the start time of
the type II (DH) radio bursts associated with the first CME shock, which occurred after the peak time of the first flare
but prior to that of the second flare. This means both energetic protons and electrons in this SEP event were accelerated
only by the first CME-driven shock. On the contrary, if we only consider the release time deduced from observations by single
spacecraft that are not well-connected (e.g. SOHO or STA in this event) via the VDA method, then, for either proton or electron,
we will mistakenly conclude that the second solar eruption is responsible for this SEP event.

 When applying the VDA analysis and in particular examining data from multiple spacecraft, we note that certain issues need to
be carefully considered. These are,

(1) In the first VDA practice mentioned in section~\ref{sect:intro}, the assumption of all particles released at the same time
is usually not satisfied (see four classes in \citet{Kim.etal14}), because the particle acceleration process is different
from event to event. Even in a single event, this condition may not be met in different longitudes, especially for
 not-well-connected points, such as the cases of SOHO and STA observations in this event.

(2) If some magnetic structures (e.g. ICME) lie between the detector and source region, the path length may have large
difference from the nominal Parker spiral path length, such as the results of the STB in this event. And the path length of
protons and electron may also differ.

(3) If particles are not released at the same time, one can use the second VDA practice mentioned above and examine
the associations and acceleration process of one SEP event \citep{Kim.etal14}. Under such cases, the assumption of the path
length being that from the nominal Parker spiral is often made. However, this is only appropriate if there is no large
magnetic structure in the IP space. For obervations made at not-well-connected vantage points, cautions must also be exercised
because time intensity profiles made at these locations may not yield clear onset times.

As a result, we suggested that
to accurately estimate the reasonable release time and path length of energetic particles via the VDA method,
the VDA should be performed using the data from a well-connected spacecraft.
This conclusion agrees with the modeling result of \citet{Wang.Qin15}.
Furthermore, the standard assumption of a scatter-free IP space needs to be made.
Under these conditions, the VDA method can be exercised in two different cases:
in the first,  one can use the VDA assuming particles of different energies are released at the same time; this practice is
often done when we are almost certain that the IP fields are not Parker-like (e.g. pre-existing MC structure between the Sun and
the spacecraft); in the second, one often adopt a nominal Parker field and then the VDA can be used to obtain the release times
for particles of different energies.

\begin{acknowledgements}
We are grateful to STEREO, SOHO, Wind, and CDAW database for making their data available online.
This work is supported at NUIST by NSFC-41304150 for Ding L.G.;
at CMA by NSFC-41274193, 41474166 for Le G.M.
\end{acknowledgements}

\bibliographystyle{raa}

%
%

\label{lastpage}

\end{document}